\documentclass[11pt]{article}

\usepackage[letterpaper, hmargin=1in, vmargin=1.25in]{geometry}

    \renewcommand{\baselinestretch}{1.4}
  \renewcommand{\arraystretch}{1.1}
\begin{document}

 \title{Improvement Of Barreto-Voloch Algorithm For Computing $r$th Roots Over Finite Fields}

 \author{Zhengjun Cao\,$^*$, \quad  Xiao Fan\\
  Department of Mathematics, Shanghai University, Shanghai,
  China.\\
  \textsf{$^*$\ \textsf{caozhj@shu.edu.cn}}  }

 \date{}\maketitle

\begin{abstract}
Root
extraction is a classical problem in computers algebra. It plays an
essential role in cryptosystems based on elliptic curves. In 2006,
Barreto and Voloch proposed an algorithm to compute $r$th roots in
${F}_{q^m} $ for certain choices of $m$ and $q$.  If
$r\,||\,q-1$ and $ (m, r)=1, $ they proved that the complexity of
their method is $\widetilde{\mathcal {O}}(r(\log m+\log\log q)m\log
q) $. In this paper, we  extend the Barreto-Voloch algorithm to the
general case that $r\,||\,q^m-1$, without the restrictions
$r\,||\,q-1$ and $(m, r)=1 $. We also specify the conditions that
the Barreto-Voloch algorithm can be preferably applied.

\end{abstract}

\section{Introduction}
Consider the problem to find a solution to  $X^r= \delta  $ in
${F}_{q^m} $, where $q=p^d$ for some prime $p$ and some integer
$d>0$. Clearly, it suffices to consider the following two cases:
$$(1)\ (r, q^m-1)=1, \qquad  (2)\ r| q^m-1 $$

 Root
extraction is a classical problem in computational algebra and
number theory. It plays an essential role in cryptosystems based on
elliptic curves. The typical applications of root extraction are
point compression in elliptic curves and operation of hashing onto
elliptic curves \cite{BBS04,BF03,Smart02}.

 Adleman,
Manders and Miller  \cite{AMM77} proposed a method to solve the
problem, which extends Tonelli-Shanks \cite{S72,T91} square root
algorithm. The basic idea of Adleman-Manders-Miller $r$th root extraction
 in ${F}_{q} $ can be described as follows. If $r|q-1$, we write $p-1$ in the form $r^{t}\cdot s$, where $(s, r)=1$.
 Given a $r$th residue $\delta$, we have
$\left(\delta^{s}\right)^{r^{t-1}}= 1   $. Since $(s, r)=1$,  it is
easy to find the least nonnegative integer $\alpha $ such that
$s|r\alpha -1$. Hence,
$\left(\delta^{r\alpha -1}\right)^{r^{t-1}}= 1  $.
If $t-1=0$, then $\delta^{\alpha }$ is a $r$th root of $\delta$.
From now on, we assume that $t \geq 2$.
 Given a $r$th non-residue $\rho\in {F}_{q}$,  we have
 $$\left(\rho^s\right)^{i\cdot r^{t-1}} \neq \left(\rho^s\right)^{j\cdot r^{t-1}}
 \ \mbox{where} \ i\neq j,\  i, j\in \{0, 1, \cdots, r-1\} $$
 Set
$K_i=\left(\rho^s\right)^{i\cdot r^{t-1}}  $ and ${K}=\{K_0,
K_1, \cdots,   K_{r-1}\}$. It is easy to find that all $K_i$ satisfy
$X^r = 1   $. Since $ \left(\left(\delta^{r\alpha
-1}\right)^{r^{t-2}}\right)^r = 1  $, there is a unique $j_1\in \{0,
1, \cdots, r-1\}$ such that
$\left(\delta^{r\alpha -1}\right)^{r^{t-2}}=K_{r-j_1}  $
(where $K_r=K_0$).  Hence,
$\left(\delta^{r\alpha -1}\right)^{r^{t-2}}K_{j_1}= 1  $. That is
$$\left(\delta^{r\alpha -1}\right)^{r^{t-2}}\left(\rho^s\right)^{j_1\cdot r^{t-1}}
= 1  $$ Likewise, there is a  unique $j_2 \in \{0, 1,
\cdots, r-1\}$ such that
$$\left(\delta^{r\alpha -1}\right)^{r^{t-3}}\left(\rho^s\right)^{j_1\cdot
r^{t-2}}\left(\rho^s\right)^{j_2\cdot r^{t-1}} = 1  \
 $$ Consequently, we obtain $j_1, \cdots, j_{t-1} $
such that
$$\left(\delta^{r\alpha -1}\right)\left(\rho^s\right)^{j_1\cdot
r}\left(\rho^s\right)^{j_2\cdot r^2}\cdots
\left(\rho^s\right)^{j_{t-1}\cdot r^{t-1}} = 1  $$
 Thus, we have
 $$\left(\delta^{\alpha}\right)^r\left(\left(\rho^s\right)^{j_1 + j_2\cdot r
 +\cdots j_{t-1}\cdot r^{t-2}}\right)^r
= \delta   $$ It means that $\delta^{\alpha}
\left(\rho^s\right)^{j_1 + j_2\cdot r
 +\cdots j_{t-1}\cdot r^{t-2}} $ is a $r$th root of
 $\delta$.  The complexity of Adleman-Manders-Miller $r$th root extraction algorithm is $\mathcal
{O}(\mbox{log}^4q+r\mbox{log}^3q)$. Notice that the algorithm  can
not run in polynomial time if $r$ is sufficiently large.

 In 2006, Barreto and Voloch \cite{BV06} proposed an
algorithm to compute $r$th roots in ${F}_{q^m} $ for certain
choices of $m$ and $q$.  If $r\,||\,q-1$ and $(m, r)=1, $ where the
notation $a^b||c$ means that $a^b$ is the highest power of $a$
dividing $c$, they proved that the complexity of their method is
$\widetilde{\mathcal {O}}(r(\log m+\log\log q)m\log q) $.

\emph{Our contributions}. We  extend the Barreto-Voloch root
extraction method to the general case that $r\,||\,q^m-1$, without
the restrictions $r\,||\,q-1$ and $(m, r)=1 $. We also specify the
conditions that the Barreto-Voloch algorithm can be preferably
applied.

\section{Barreto-Voloch method }
Barreto-Voloch method  takes advantage of the periodic structure of
$v$ written in base $q$ to compute $r$th roots in ${F}_{q^m}
$, where $v=r^{-1} \, (\mbox{mod}\, q^m-1)$ if $(r, q^m-1)=1 $. This
advantage is based on the following fact \cite{BV06}:

Fact 1. \emph{Let ${F}_{q^m} $ be a finite field of characteristic
$p$ and let $s$ be a power of $p$. Define the map
$$\phi_n :{F}_{q^m}\rightarrow {F}_{q^m},
y\mapsto y^{1+s+\cdots+s^n}\  \mbox{for}\, \ n\in {N}^* $$ We
can compute $\phi_n(y)$ with $\mathcal {O}(\log n)$ multiplications
and raisings to powers of $p$.}

Notice that   raising to powers of $p$ has negligible cost,  if
we   use a normal basis for  ${F}_{q^m}/{F}_{q} $.
Since it only requires $\mathcal {O}(\log n)$ multiplications
  and raisings to powers of $p$ to compute $y^{1+s+\cdots
+ s^n}$, where $p$ is the characteristic of ${F}_{q^m}$ and
$s$ is a power of $p$, their method becomes more efficient for certain
choices of $m$ and $q$. They obtained the following results
\cite{BV06}.

Lemma 1.  \emph{Given $q$ and $r$ with $(q(q-1), r)=1$, let $k>1$ be
the order of $q$ modulo $r$. For any $m>0$, $(m, k)=1$, let $ u,
1\leq u<r$ satisfy $u(q^m-1)\equiv -1 \, (\emph{\mbox{mod}}\, r)$
and $v = \lfloor q^mu/r\rfloor$. Then $rv \equiv 1 \,
(\emph{\mbox{mod}}\, q^m-1)$. In addition, $v =a +b \sum^{n-1}_{j=0}
q^{jk}, a, b< q^{2k}, n=\lfloor m/k\rfloor$.}

Theorem 1. \emph{Let $q$ be a prime power, let $r >1$ be such that
$(q(q-1), r)=1$ and let $k>1$ be the order of $q$ modulo $r$. For
any $m>0, (m, k)=1$, the complexity of taking $r$th roots in
${F}_{q^m} $ is $\widetilde{\mathcal {O}}((\log m+r\log
q)m\log q) $.}

Lemma 2.   \emph{Given $q$ and $r$ with $r\,|\,(q-1)$ and $((q-1)/r,
r)=1$, for any $m >0$, $(m, r)=1$, let $u, 1\leq u<r$ satisfy
$u(q^m-1)/r \equiv -1 \, (\emph{\mbox{mod}}\, r)$ and $v = \lceil
q^mu/r\rceil $. Then $ rv \equiv 1 \, (\emph{\mbox{mod}}\,
(q^{m}-1)/r^2)$. In addition, $v =a +b \sum^{n-1}_{j=0} q^{jr}, a,
b< q^{2r}, n=\lfloor m/r\rfloor$.}

Theorem 2. \emph{Let $q$ be a prime power and let $r >1$ be such
that $r\, |\, (q-1)$ and $((q-1)/r, r)=1$. For any $m>0, (m, r)=1$,
given $ x\in {F}_{q^m} $ one can compute the $r$th root of
$x$ in ${F}_{q^m}$, or show it does not exist, in
$\widetilde{\mathcal {O}}(r(\log m+\log\log q)m\log q) $ steps.}

\section{Analysis of Barreto-Voloch method }

\subsection{On the  conditions of  Barreto-Voloch method}

In  Theorem 1,  it requires that $$(q(q-1), r)=1 \  \mbox{and}\  (m,
k)=1$$ where $k>1$ is the order of $q$ modulo $r$. These conditions
imply $(q^m-1, r)=1 $. But these are not necessary to the general
case. Likewise,  in  Theorem 2, it requires that $$r\,||\,q-1\
\mbox{and} \ (m, r)=1$$  These imply $r\,||\,q^m-1 $. But these are
not necessary, too.
 We will
remove the restrictions and investigate the following cases:\\
\centerline {(1) $(r, p^m-1)=1$; \qquad   (2) $r\,||\,p^m-1 $.}
where $p$ is a prime. As for the general case, $p^m-1=r^{\alpha}s,
\alpha\geq 2, (r, s)=1 $, we refer to \cite{AMM77}.

\subsection{On the technique of periodic structure}

As we mentioned before, Barreto-Voloch method  takes advantage of
the periodic structure of $v$ written in base $q$. Precisely, in
Lemma 1
$$v =a
+b \sum^{n-1}_{j=0} q^{jk}, a, b< q^{2k}, n=\lfloor m/k\rfloor
\eqno(1) $$  where $k>1$ is the order of $q$ modulo $r$. From the
expression,  we know it requires  that  $n=\lfloor m/k\rfloor \geq 1
$. It is easy to find that \emph{the advantage of Barreto-Voloch
method due to the periodic expansion in base $q$ requires that $m$
is much greater than $k$.} That is, the length of such periodic
expansion, $n$, should be as large as possible.

Since  raising to a power of $p$ is a linear bijection in
characteristic $p$, the complexity of such operation is no larger
than that of multiplication, namely, $\widetilde{\mathcal {O}}(m\log
p) $ using FFT techniques \cite{GGPS,GG03,S05}. In light of that
$q=p^d$ for some prime $p$, it is better to write $v$ as
$$v =a' +b'
\sum^{n'-1}_{j=0} p^{jk'}, a', b'< p^{2k'}, n'=\lfloor md/k'\rfloor
\eqno(2)$$ where $k'$ is the order of $p$ modulo $r$. That is, the
periodic expansion in base $p$ could produce a large expansion
length, instead of the original periodic expansion in base $q$.
 This claim is directly based on the following  fact
   $$n'=\lfloor md/k'\rfloor \geq \lfloor md/kd\rfloor=n \eqno(3)
$$ (This is because  $k'\,|\,kd$. See the definitions of $k, k'$.)

\section{Extension of Barreto-Voloch method }

\subsection{Taking $r$th roots when $r$ is invertible}

We first discuss the problem to take $r$th roots over
${F}_{p^m} $ if $(r, p^m-1)=1$, where $p$ is a prime.

   Lemma 3.  \emph{Suppose that
$(p^m-1, r)=1$. Let $k$ be the order of $p$ modulo $r$. Let $ u,
1\leq u<r$ satisfy $u(p^m-1)\equiv -1 \, (\emph{\mbox{mod}}\, r)$.
 Then $rv \equiv 1 \,
(\emph{\mbox{mod}}\, p^m-1)$, where $v = \lfloor p^mu/r\rfloor$. In
addition, if $m>k$, then $v =a +b \sum^{n-1}_{j=0} p^{jk}, a, b<
p^{2k}, n=\lfloor m/k\rfloor$.}

\emph{Proof}. Since $u(p^m-1)\equiv -1 \, ({\mbox{mod}}\, r)$ and
$1\leq u<r$, we have  $p^mu/r=\lfloor p^mu/r\rfloor+ (u-1)/r$ and
$r\lfloor p^mu/r\rfloor \equiv 1 \, ({\mbox{mod}}\, p^m-1)$. Let
$z=u(p^k-1)/r$. Then $z$ is an integer and $z<p^k-1$. Hence,
$p^mu/r=p^mz/(p^k-1)  $. If $m>k$, then we have the following
expansion
$$p^mz/(p^k-1)=p^{m-k}z\sum_{j=0}^{\infty}p^{-jk}=
p^{m-nk}z\sum_{j=0}^{n-1}p^{jk}+ p^{m-k}z\sum^{\infty}_{n}p^{-jk}
$$ Take $a=\lfloor p^{m-k}z\sum^{\infty}_{n}p^{-jk}\rfloor,
b=p^{m-nk}z$. This completes the proof. 

Theorem 3. \emph{Suppose that $(p^m-1, r)=1$. Let $k$ be the order
of $p$ modulo $r$. If $m>k$, then the complexity of taking $r$th
roots of $\delta$ in ${F}_{p^m} $ is $\widetilde{\mathcal
{O}}((\log m+k\log p)m\log p) $.}

\emph{Proof}. Given $\delta\in {F}_{p^m} $, clearly,
$\delta^{r^{-1}}$ is a root of $X^r= \delta $
 if $(p^m-1, r)=1$, where $r^{-1}$ is the inverse of $r$ modulo $p^m-1$.

 By Lemma 3, if $m>k$, then $r^{-1} =a +b
\sum^{n-1}_{j=0} p^{jk} \, ({\mbox{mod}}\, p^m-1),  a, b< p^{2k},
n=\lfloor m/k\rfloor$. Raising to the power $\sum^{n-1}_{j=0}
p^{jk}$ takes $\mathcal {O}(\log n)$  multiplications and raisings
to powers of $p$. The raising to the power $a$  takes $\mathcal
{O}(k \log p)$ multiplications due to the bound on the exponent. So
does the raising to the power $b$. The total computation cost is
therefore $\mathcal {O}(\log m+ k \log p)$ operations of complexity
$\widetilde{\mathcal {O}}(m\log p) $ (if directly using the form
$r^{-1}=\frac{u(p^m-1)+1}{r} $, it takes time $\widetilde{\mathcal
{O}}(m^2\log^2 p) $). This completes the proof.

\subsection{Taking $r$th roots when $r$ is not invertible}

We now discuss the problem to take $r$th roots over
${F}_{p^m} $ if $r\,||\,p^m-1$, where $p$ is a prime.

Lemma 4.   \emph{Suppose that $r\,||\,p^m-1$. Let $k$ be the order
of $p$ modulo $r$. Let $u, 1\leq u<r$ satisfy $u(p^m-1)/r \equiv -1
\, (\emph{\mbox{mod}}\, r)$ and $v = \lceil p^mu/r^2\rceil $. Then $
rv \equiv 1 \, (\emph{\mbox{mod}}\, (p^{m}-1)/r)$. In addition, if
$m>kr$, then $v =a +b \sum^{n-1}_{j=0} p^{jkr}, a, b< p^{2kr},
n=\lfloor m/kr\rfloor$.}

\emph{Proof}. Since $u(p^m-1)/r\equiv -1 \, ({\mbox{mod}}\, r)$ and
$1\leq u<r$, we have  $p^mu/r^2=\lceil p^mu/r^2\rceil+ (u-r)/r^2$
and $r\lceil p^mu/r^2\rceil \equiv 1 \, ({\mbox{mod}}\, (p^m-1)/r)$.
Let $z=u(p^{kr}-1)/r^2$. Then $z$ is an integer and $z<p^{kr}-1$.
Hence, $p^mu/r^2=p^mz/(p^{kr}-1)  $. If $m>kr$, then we have the
following expansion
$$p^mz/(p^{kr}-1)=p^{m-kr}z\sum_{j=0}^{\infty}p^{-jkr}=
p^{m-nkr}z\sum_{j=0}^{n-1}p^{jkr}+
p^{m-kr}z\sum^{\infty}_{n}p^{-jkr}
$$ Take $a=\lfloor p^{m-kr}z\sum^{\infty}_{n}p^{-jkr}\rfloor,
b=p^{m-nkr}z$. This completes the proof.

Theorem 4. \emph{Suppose that $r\,||\,p^m-1$. Let $k$ be the order
of $p$ modulo $r$.
   If $m>kr$, then one can compute the $r$th
root of $\delta$ in ${F}_{p^m}$, or show it does not exist,
in $\widetilde{\mathcal {O}}((\log m+ kr \log p)m\log p) $ steps.}

\emph{Proof}. Given $\delta\in {F}_{p^m} $, we have
$\delta^{p^m-1}=1$.  If $r\,||\,p^m-1$ and $\delta^{(p^m-1)/r}=1$,
then there exists an integer $v$ such that $\frac{p^m-1}r\,|\,vr-1$
and $(\delta^v)^r=\delta$. Hence, it suffices to compute the inverse
of $r$  modulo $\frac{p^m-1}r$.

By Lemma 4, if $m>kr$, $r^{-1}\equiv v =a +b \sum^{n-1}_{j=0}
p^{jkr} \, ({\mbox{mod}}\, (p^m-1)/r),  a, b< p^{2kr}, n=\lfloor
m/kr\rfloor$,
 Since raising to the power $\sum^{n-1}_{j=0} p^{jkr}$
takes $\mathcal {O}(\log n)$  multiplications and raisings to powers
of $p$. Raising to the power $a$  takes $\mathcal {O}(kr \log
p)$ multiplications due to the bound on the exponent. So does
raising to the power $b$.
 The cost of
raising to $v$ is therefore $\mathcal {O}(\log m+ kr \log p)$
operations of complexity $\widetilde{\mathcal {O}}(m\log p) $. To
check that $\rho =\delta^v$ is a correct root,  we compute $\rho^r$
with cost $\widetilde{\mathcal {O}}(m\log r \log p) $. If $\delta$
is a $r$th power, then  $\rho^r =\delta$, otherwise $\rho^r$ is not
equal to $\delta$. The total computation cost is therefore
$\widetilde{\mathcal {O}}((\log m+ kr \log p)m\log p) $ (if directly
using the form $r^{-1}=\frac{u(p^m-1)+r}{r^2} $, it takes time
$\widetilde{\mathcal {O}}(m^2\log^2 p) $). This completes the proof.

\section{Conclusion}

In this paper, we analyze and extend the Barreto-Voloch method to
compute $r$th roots over finite fields.    We specify the conditions
that the Barreto-Voloch algorithm can be preferably applied. We also
give a formal complexity analysis of the method.

 \emph{Acknowledgements}
  This work is supported by the National Natural Science
Foundation of China (Project 60873227), and the Key Disciplines of
 Shanghai Municipality (S30104).

\end{document}